\documentclass [preprint, amsmath,amssymb,aps,prb]{revtex4-1}

\usepackage{graphicx}
\usepackage{amsmath}
\usepackage{amssymb}

\begin{document}

\title{Tunable band structure and effective mass of disordered chalcopyrite } 

\author{Wang Ze-Lian}
 \affiliation{College of Physics and Electronic Engineering,
              Institute of Solid State Physics,
              Sichuan Normal University, Chengdu 610068, P. R. China}

\author{Xie Wen-Hui}
 \affiliation{Department of Physics, 
              East China Normal University, Shanghai 200062, P. R. China}

\author{Zhao Yong-Hong}
 \email{yhzhao@sicnu.edu.cn}
 \affiliation{College of Physics and Electronic Engineering,
              Institute of Solid State Physics,
              Sichuan Normal University, Chengdu 610068, P. R. China}

\begin{abstract}
The band structure and effective mass of disordered chalcopyrite photovoltaic materials Cu$_{1-x}$Ag$_x$Ga$X_2$ ($X=$ S, Se) are investigated by density functional theory. Special quasirandom structures are used to mimic local atomic disorders at Cu/Ag sites. A local density plus correction method is adopted to obtain correct semiconductor band gaps for all compounds. The bandgap anomaly can be seen for both sulfides and selenides, where the gap values of Ag compounds are larger than those of Cu compounds. Band gaps can be modulated from 1.63 to 1.78 eV for Cu$_{1-x}$Ag$_x$Ga$Se_2$, and from 2.33 to 2.64 eV for Cu$_{1-x}$Ag$_x$Ga$S_2$. The band gap minima and maxima occur at around $x=0.5$ and $x=1$, respectively, for both sulfides and selenides. In order to show the transport properties of Cu$_{1-x}$Ag$_x$Ga$X_2$, the effective mass is shown as a function of disordered Ag concentration. Finally, detailed band structures are shown to clarify the phonon momentum needed by the fundamental indirect-gap transitions. These results should be helpful in designing high-efficiency photovoltaic devices, with both better absorption and high mobility, by Ag-doping in CuGa$X_2$.
\end{abstract}
\maketitle

Although new photovoltaic materials, such as organic perovskites, have attracted much attention recently,\cite{perov1,perov2,perov3} zinc-blende semiconductors still play an important role, owing to their high efficiency and stability in real environments. For example, Cu$_2$ZnSnSe$_4$ \cite{AM2013} (CZTS) and Cu(In,Ga)Se$_2$ \cite{JACS2008} (CIGS), which are both based on chalcopyrite CuGa$X_2$ ($X=$ S, Se), have been considered as potential candidates for photovoltaics. It has been reported recently that the efficiency of CIGS has gone beyond 20\%, which is close to that of polycrystalline silicon.\cite{efficiency2014} Furthermore, photovoltaic devices related to CZTS have gained attention because they solely contain abundant and nontoxic materials.\cite{AM2013,CZTS}

The spectrum of sunshine arriving at the earth ranges from 0.35 to 2500 nm.\cite{spectrum2011} Therefore, tunable bandgaps are essential for photovoltaic materials, in order to absorb solar flux as much as possible. The use of doping impurities is the most commonly used method to modulate the bandgaps, as well as effective mass (EM), of semiconductors.\cite{tune1988, tune2002, EM1961} Phase stability and formation energy of uniform impurities in CZTS have been systematically reviewed and classified using the supercell method.\cite{AM2013,PRB2013} Owing to high manufacturing costs, practical photovoltaic devices cannot be made of crystalline CZTS, and hence atomic disordered defects and impurities are of high importance in practical devices. Although successful in uniform situations, the supercell method is difficult to apply to highly disordered impurities, which are very common in practical photovoltaic devices. In addition, the energy band structure is a key factor for both optical and electronic transport properties of photovoltaic materials. However, there is a well-known underestimation of semiconductor bandgaps by first-principles calculations. There are several common methods to modify the bandgap underestimation of density functional theory, such as the Heyd--Scuseria--Ernzerhof (HSE) hybrid functional \cite{HSE}, quasiparticle approximation with Green's Function (GW) \cite{GW}, and the modified Becke--Johnson (mBJ) \cite{mBJ} method. The HSE and GW calculations are extremely expensive in calculation, while the mBJ method can only give a partially corrected bandgap of 1.03 eV for CuGaSe$_2$\cite{mBJCuGaSe}, which is still much smaller than the experimental value of 1.68 eV. Therefore, it is desirable to find a first-principles scheme to overcome the well-known bandgap underestimation for semiconductors, which is also capable of treating disordered impurities including spin-orbit coupling (SOC), with moderate computational cost.

    \begin{table}[htbp]
     \caption{\label{tab:tab1} Full optimized lattice constants (a/c) of disordered Cu$_{1-x}$Ag$_x$Ga$X_2$ ($X=$ Se, S) given by SQS method in unit of \AA.}
    \begin{ruledtabular}
    \begin{tabular}{cccc}
    &$x$    &Cu$_{1-x}$Ag$_x$GaSe$_2$  &Cu$_{1-x}$Ag$_x$GaS$_2$  \\
    &0.00   &11.05/10.97     &10.48/10.42        \\
    &0.25   &11.18/11.05     &10.62/10.48        \\
    &0.50   &11.36/11.09     &10.82/10.51        \\
    &0.75   &11.50/11.08     &10.97/10.52        \\
    &1.00   &11.70/11.05     &11.17/10.51        \\
    \end{tabular}
    \end{ruledtabular}
    \end{table}

Although CZTS and CIGS are the most important chalcopyrite photovoltaic materials, Cu$_{1-x}$Ag$_x$Ga$X_2$ ($X=$ Se, S) have also drawn much attention, owing to their interesting band anomalies and grain boundary effects.\cite{anormal,SQS1,grain2015} In this work, we focus on the band structure and EM of Cu$_{1-x}$Ag$_x$Ga$X_2$, modulated by disordered Ag atoms. The special quasirandom structure (SQS) method has been proved reasonable for disorder effects in all these compounds with $x$ ranging from 0.0 to 1.0.\cite{SQS1,SQS2,SQS3} The initial guesses of internal parameters are given by Ref. \onlinecite{SQS1}. All electronic structure calculations are based on fully optimized SQS structures, including both lattice constants and internal parameters. The Vienna \textit{ab-initio} simulation package is used for optimization, with the local density approximation (LDA) for exchange-correlation potential and projector augmented wave for pseudopotentials.\cite{vasp,lda,paw} The kinetic energy cutoff for a plane wave basis is set to 800 eV and a $k$-mesh of $4\times 4 \times 4$ is used. The optimized lattice constants are consistent with previous results for pure CuGa$X_2$\cite{latticeSe,latticeS} and AgGa$X_2$\cite{latticeAg}, as listed in TABLE \ref{tab:tab1}. It can be seen that sulfides have smaller lattice constants than selenides. For both sulfides and selenides, $a$ increases with increasing Ag concentration, while there is a maximum value of $c$ at approximately $x=0.75$ and 0.5 for the sulfides and selenides, respectively. This indicates that there is a compression strain along the $z$-direction for all compounds when the Ag concentration is larger than the threshold value.

\begin{figure}[htbp]
 \includegraphics[width=8cm]{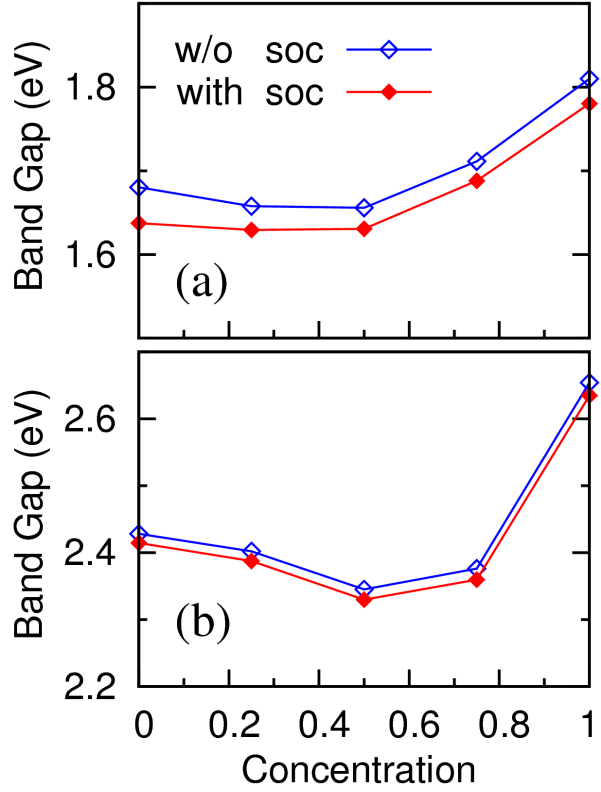}
 \caption{\label{fig:fig1} Bandgaps as a function of Ag concentration, (a) for Cu$_{1-x}$Ag$_x$GaSe$_2$ and (b) for Cu$_{1-x}$Ag$_x$GaS$_2$. The blue hollow diamonds show that without SOC, while the red solid diamonds show that with SOC included.}
\end{figure}

The electronic structure calculations are based on the linear combination of atomic orbital (LCAO) method as implemented in Nanodcal package.\cite{nanodcal,nanodcal1,nanodcal2} A set of optimized double-$\zeta$ polarization (DZP) LCAO basis is used for all atoms, together with a real space grid energy cutoff of 300 Ry and a $k$-grid of $4 \times 4 \times 4$ in the first Brillouin zone. Self-consistent calculations are considered to be converged when the charge density has converged to less than 10$^{-5}$. In this work, we use a LDA plus correction method (LDA+C) \cite{LDAC} to modify the bandgaps of Cu$_{1-x}$Ag$_x$Ga$X_2$. It is shown that experimental gap values can be achieved by choosing appropriate orbit-dependent correction parameters for each element, as given in TABLE \ref{tab:tab2}. The calculated bandgaps for Cu$_{1-x}$Ag$_x$GaSe$_2$ and Cu$_{1-x}$Ag$_x$GaS$_2$ for $x=$ 0.0, 0.25, 0.5, 0.75, and 1.0 are shown in FIG. \ref{fig:fig1}. Spin-orbit interaction is important for the electronic structure of semiconductors, which is included in all calculations. The solid red diamonds show the results including SOC, while the hollow blue diamonds show those without SOC. It can be seen that there is a gap minimum for Ag concentration near 0.5 for both sulfides and selenides. The results show that SOC interaction can decrease the bandgaps significantly for selenides, but only modestly for sulfides. The calculated bandgaps are consistent with experimental\cite{gapexp} and semiclassical\cite{SQS1} values.

\begin{table}
   \caption{\label{tab:tab2} Orbit-dependent correction values in LDA+C calculation for all elements in Cu$_{1-x}$Ag$_x$Ga$X_2$.}
    \begin{ruledtabular}
    \begin{tabular}{ccccc}
           &Element   &s   &p   &d\\
           &Cu   &0.00   &0.00   &4.73\\
           &Ag   &0.00   &0.00   &5.46\\
           &Ga   &3.00   &0.00   &0.00\\
           &Se   &3.00   &0.00   &0.00\\
           &S    &3.90   &0.28   &9.40
    \end{tabular}
    \end{ruledtabular}
\end{table}

\begin{figure}[htbp]
 \includegraphics[width=8cm]{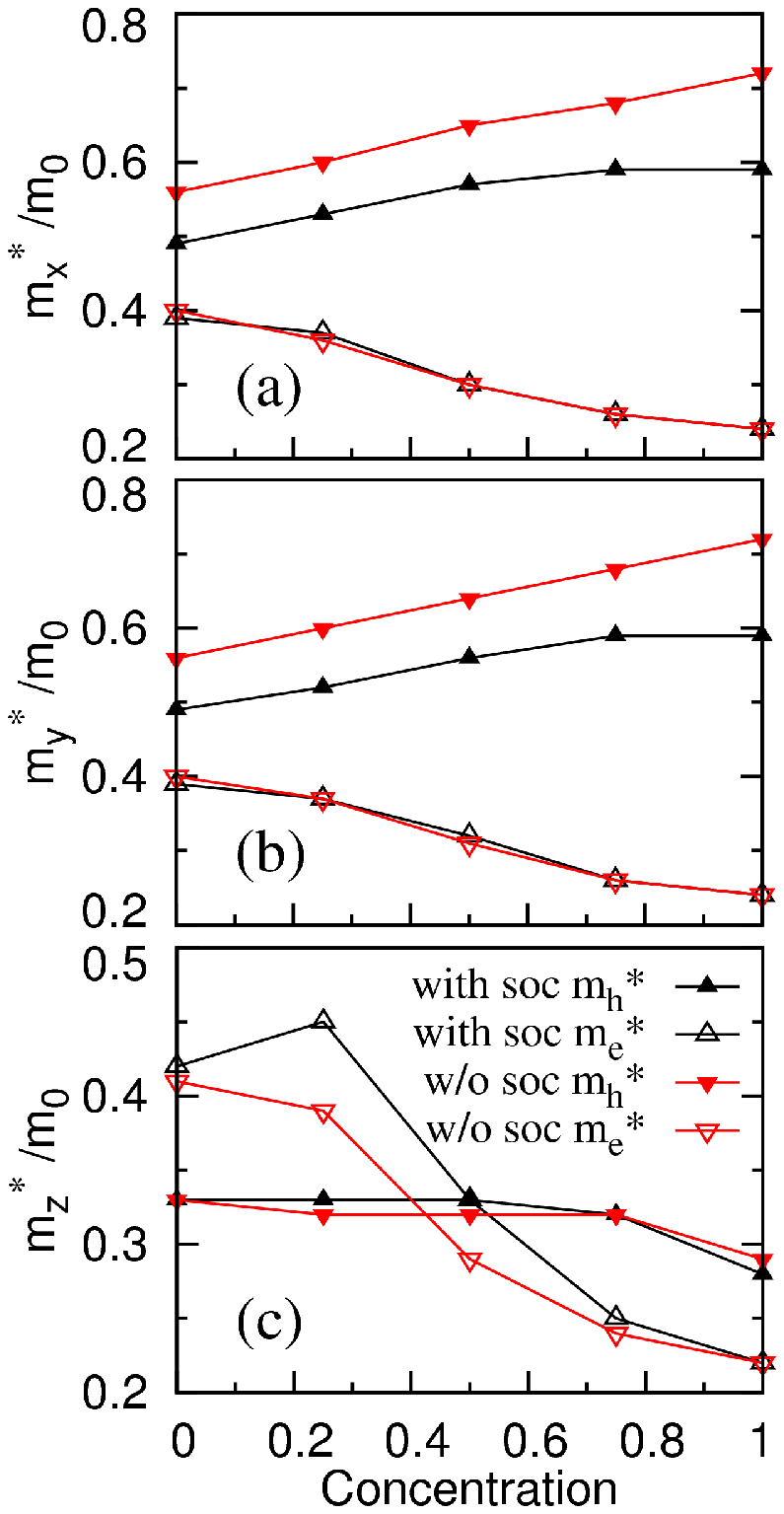} 
 \caption{\label{fig:fig2} Anisotropic EM of Cu$_{1-x}$Ag$_x$GaSe$_2$, where (a), (b), and (c) show the $x$-, $y$- and $z$-directions, respectively. In each subfigure, the solid and/or hollow triangles show hole and/or electron EM, respectively. The upward and downward triangles show those with and without SOC.}
\end{figure}

EM is important for the electronic transport properties in photovoltaic materials. Anisotropic EM is defined as follows,
 \begin{equation}
   m_i^* = \frac{\hbar^2}{d^2E/dk_i^2}, (i=x,y,z).
 \end{equation} 
The calculated EM of Cu$_{1-x}$Ag$_x$GaSe$_2$ is shown in FIG. \ref{fig:fig2}. These results indicate that EM behaves similarly in the $k_x$- and $k_y$-directions, where the hole EM increases with Ag concentration and the electron EM decreases with it. Moreover, SOC can decrease the hole EM to approximately 0.07 m$_0$, but keep the electron EM unchanged. Owing to the fact that the optimized lattice constants of SQSs deviate significantly from the cubic structure as given in TABLE \ref{tab:tab1}, EM in the $k_z$-direction differs significantly from those in the $k_x$- and $k_y$-directions. The hole EM in the $k_z$-direction remains nearly constant for Ag concentration ranging from 0.0 to 0.75, and then decreases slightly until $x=1.0$. However, the electron EM decreases remarkably with increasing Ag concentration. For Cu$_{1-x}$Ag$_x$GaS$_2$, there is a maximum at $x=0.25$ for electron EM and SOC has little effect on EM in the $k_x$- and $k_y$-directions. EM in the $k_z$-direction behaves similar to that of Cu$_{1-x}$Ag$_x$GaSe$_2$. 

\begin{figure}[htbp]
 \includegraphics[width=8cm]{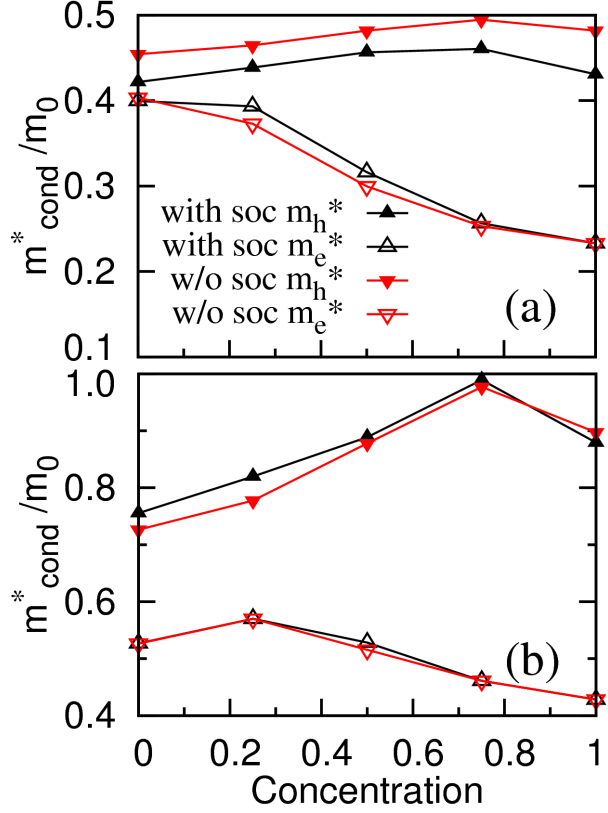}
 \caption{\label{fig:fig3} EM for conductivity of (a) Cu$_{1-x}$Ag$_x$GaSe$_2$ and (b) Cu$_{1-x}$Ag$_x$GaS$_2$, where the upward and downward solid triangles show hole EM for with and without SOC, and hollow triangles for electron EM.}
\end{figure}

There are several type of EMs for different purposes, such as EM for density of states and conductivity, among others. For photovoltaic applications, EM for conductivity is of most importance, which is related to photocurrent and defined as follows,
 \begin{equation}
    m^*_{cond} = 3 [\frac{1}{m_x^*} + \frac{1}{m_y^*} + \frac{1}{m_z^*}]^{-1}.
 \end{equation} 
The calculated $m^*_{cond}$ as a function of Ag concentration is shown in FIG. \ref{fig:fig3}. FIG. \ref{fig:fig3} (a) shows $m^*_{cond}$ of Cu$_{1-x}$Ag$_x$GaSe$_2$. The hole EM increases slowly from 0.42$m_0$ and reaches a maximum of 0.46$m_0$ at approximately $x=0.75$, while the electron EM quickly decreases from 0.40$m_0$ to 0.23$m_0$ with Ag concentration increasing from 0.0 to 1.0. $m^*_{cond}$ of Cu$_{1-x}$Ag$_x$GaS$_2$ is shown in FIG. \ref{fig:fig3} (b). The hole EM increases from 0.76$m_0$ and reaches the maximum of 0.99$m_0$ at approximately $x=0.75$. The electron EM begins with 0.53$m_0$ at $x=0.0$, then reaches a maximum of 0.57$m_0$ at approximately $x=0.25$ and a minimum of 0.43$m_0$ at $x=1.0$. Generally speaking, a smaller EM implies a higher carrier mobility. Therefore, the electron mobility of Cu$_{1-x}$Ag$_x$GaSe$_2$ can be promoted by increasing Ag concentration, while accompanied by some reduction of hole mobility. For Cu$_{1-x}$Ag$_x$GaS$_2$, the electron mobility can be promoted only when the Ag concentration is larger than 0.25, and accompanied by a large reduction of hole mobility. On balance, the mobility of selenides is higher than that of sulfides for both electrons and holes, and should be enhanced by increased Ag concentration.

\begin{figure}[htbp]
 \includegraphics[width=8cm]{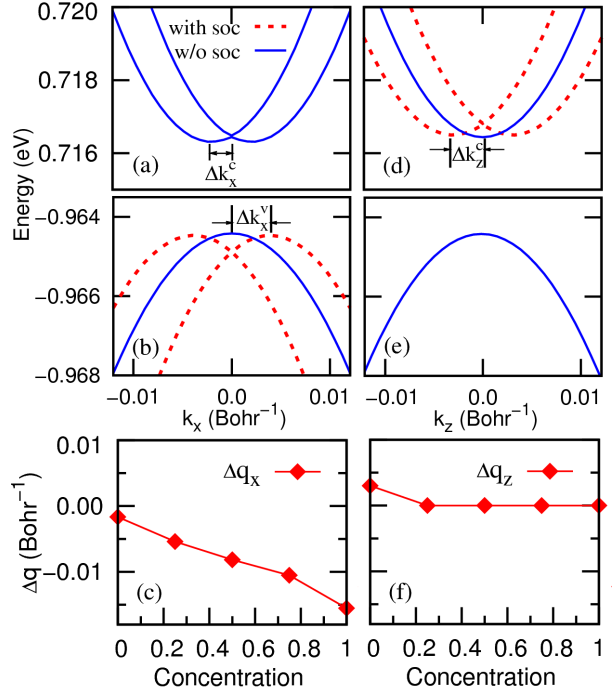}
 \caption{\label{fig:fig4} Anisotropic band structure of Cu$_{1-x}$Ag$_x$GaSe$_2$, where (a) and (b) for conduction bands and valence bands in the $k_x-$direction, respectively. The blue lines show bands without SOC, while the red dot line those including SOC. (d) and (e) show band structure in $k_z-$direction. The momentum of phonon needed by fundamental transitions in the $k_x-$ and $k_z-$directions are presented in (c) and (f). }
\end{figure}

Optic absorption is one of the most important properties of photovoltaic materials. The optical transitions between any bands must obey conservation of both energy and momentum. Indirect-gap semiconductors undergo a phonon-assisted transition. However, for fundamental transition of direct-gap semiconductors, phonons are unnecessary, which typically results in higher absorption. On the other hand, direct bandgap will unfortunately result in short minority carrier lifetime. The detailed band structure of Cu$_{1-x}$Ag$_x$GaSe$_2$ is shown in FIG. \ref{fig:fig4}, where $\Delta k_i^C, (i=x,y,z)$ shows the deviation of the conduction band minimum (CBM) and $\Delta k_i^V$ indicates the deviation of the valence band maximum (VBM) from the $\Gamma$ point. $\Delta q_i = \Delta k_i^C - \Delta k_i^V$ is the phonon momentum needed by the fundamental indirect-gap transition. It can be seen from FIG. \ref{fig:fig4}(a) and (b) that SOC causes the top valence band in the $k_x$-direction of CuGaSe$_2$ to split about 0.005 Bohr$^{-1}$, but leaves the bottom conduction band unsplit. FIG. \ref{fig:fig4}(c) shows $\Delta q_x$ as a function of Ag concentration, which indicates that a larger phonon momentum is needed for larger Ag concentration compounds. The band structure in the $k_y$-direction is absolutely similar to that in the $k_x$-direction. The CBM and VBM of CuGaSe$_2$ in the $k_z$-direction are different and shown in FIG. \ref{fig:fig4}(d) and (e), where it can be seen that VBM remains in the $\Gamma$ point, regardless of SOC or an increase in Ag concentration. When there is no Ag, SOC causes a splitting of CBM, which disappears quickly with increasing Ag concentration. This means that a reduction of Ag ions can lead to a direct bandgap in the $k_z$-direction of Cu$_{1-x}$Ag$_x$GaSe$_2$.

In summary, we calculate the electronic band structure and effective mass of disordered quaternary chalcopyrite Cu$_{1-x}$Ag$_x$Ga$X_2$ ($X=$ Se, S) as a function of Ag concentration. Disorder structures are depicted by SQS method and a LDA+C potential is used to correct the well-known bandgap underestimation of the first-principles calculation. SOC is included for all compounds. For both sulfides and selenides, the bandgap minimum appears at 50\% Ag concentration. Meanwhile, bandgaps of AgGa$X_2$ are bigger than CuGa$X_2$, which is consistent with the abnormal band structure of the previous study. The anisotropic EMs of Cu$_{1-x}$Ag$_x$Ga$X_2$ have been calculated, from which the conductivity related to EM can be derived. The results show that the electron EM of Cu$_{1-x}$Ag$_x$Ga$Se_2$ decreases with increasing Ag concentration, and the hole EM remains nearly constant. However, there are clear maxima at $x=0.25$ and 0.75 for the electron and hole EM, respectively, of Cu$_{1-x}$Ag$_x$Ga$S_2$. Moreover, the phonon momentum needed by fundamental indirect-gap transition will be increased substantially in the $k_x-$ and $k_y-$directions by increased Ag concentration. Therefore, absorption along the $k_x-$ and $k_y-$directions can be reduced by increased Ag concentration. In contrast, the bandgap in the $k_z-$direction becomes direct and the corresponding phonon momentum quickly goes to zero as the Ag concentration approaches 25\%. Therefore, we deduce that absorption will be much more efficient in the $k_z-$direction than in the $k_x-$ and $k_y-$directions. Our calculated results will be useful in designing high-performance photovoltaic devices.

We thank National Natural Science Fundation of China (No. 11104191 and 51572086) and Sichuan Normal University for financial support.


\begin{thebibliography}{99} 

\bibitem{perov1} N. J. Jeon, J. H. Noh, Y. C. Kim, W. S. Yang, S. Ryu, and S. Seok, Nat. Mater. {\bf 13}, 897 (2014).

\bibitem{perov2} M. Liu, M. B. Johnston, and H. J. Snaith, Nature {\bf 501}, 395 (2013).

\bibitem{perov3} H. Zhou, Q. Chen, G. Li, S. Luo, T. Song, H. Duan, Z. Hong, H. You, Y. Liu, and Y. Yang, Science {\bf 345}, 542 (2014).

\bibitem{AM2013} S. Chen, A. Walsh, X. G. Gong, and S. H. Wei, Adv. Mater. {\bf 25}, 1522 (2013).

\bibitem{JACS2008} M. G. Panthani, V. Akhanvan, B. Goodfellow, J. P. Schmidtke, L. Dunn, A. Dodabalapur, P. F. Barbara, and B. A. Korgel, J. Am. Chem. Soc. {\bf 130}, 16770 (2008).

\bibitem{efficiency2014} M. A. Green, K. Emery, Y. Hishihiro, W. Warta, and E. D. Dunlop, Prog. Photon.: Res. Appl. {\bf 23}, 1 (2014).

\bibitem{CZTS} S. Abermann, Solar Energy {\bf 94}, 37 (2013).

\bibitem{spectrum2011} T. P. Otanicar, I. Chowdhury, R. Prasher, and P. E. Phelan, J. Sol. Energy Eng. {\bf 133}, 041014 (2011).

\bibitem{tune1988} B. E. Sernelius, K. -F. Berggren, Z. -C. Jin, I. Hamberg, and C. G. Granqvist, Phys. Rev. B {\bf 37}, 10244 (1988).

\bibitem{tune2002} X. Nie, S. H. Wei, S. B. Zhang, Phys. Rev. Lett. {\bf 88}, 066405 (2002).

\bibitem{EM1961} M. Cardona, Phys. Rev. {\bf 121}, 752 (1961).

\bibitem{PRB2013} J. Pohl and K. Albe, Phys. Rev. B {\bf 87}, 245203 (2013).

\bibitem{HSE} A. V. Krukau , O. A. Vydrov, A. F. Izmaylov, and G. E. Scuseria, J. Chem. Phys. {\bf 125}, 224106 (2006).

\bibitem{GW} M. S. Hybertsen and S. G. Louie, Phys. Rev. B, {\bf 34}, 5390 (1986).

\bibitem{mBJ} F. Tran and P. Blaha, Phys. Rev. Lett., {\bf 102}, 226401 (2009).

\bibitem{mBJCuGaSe} J. Srour, M. Badawi, F. EI H. Hassan, and A. V. Postnikov, Phys. Status Solidi B, {\bf 253}, 1472 (2016).

\bibitem{anormal} $Semiconductors: Data Handbook$, 3rd ed., edited by O. Madelung (Springer, Berlin, 2014).

\bibitem{SQS1}  S. Chen, X. G. Gong, and S. H. Wei, Phys. Rev. B {\bf 75}, 205209 (2007).

\bibitem{grain2015} H. Mirhosseini, H. Kiss, and C. Felser, Phys. Rev. Appl. {\bf 4}, 064005 (2015).

\bibitem{SQS2} A. Zunger, S. H. Wei, L. G. Ferreira, and J. E. Bernard, Phys. Rev. Lett. {\bf 65}, 353 (1990).

\bibitem{SQS3} S. H. Wei, L. G. Ferreira, J. E. Bernard, and A. Zunger, Phys. Rev. B {\bf 42}, 9622 (1990).

\bibitem{vasp} G. Kresse and J. Furthm{\"u}ller, Phys. Rev. B {\bf 54}, 11169 (1996).

\bibitem{lda} D. M. Ceperley and B. J. Alder, Phys. Rev. B {\bf 45}, 566 (1980).

\bibitem{paw} P. E. Bl{\"o}chl, Phys. Rev. B {\bf 50}, 17953 (1994).

\bibitem{latticeSe} T. Maeda and T. Wada, J. Phys. Chem. Sol. {\bf 66}, 1924 (2005).

\bibitem{latticeS} G. D. Body, H. Kasper, and J. H. Mcfee, IEEE J. Quan. Elec. {\bf 7}, 563 (1971).

\bibitem{latticeAg} B. Tell and H. M. Kasper, Phys. Rev. B {\bf 4}, 4455 (1971).

\bibitem{nanodcal} J. Taylor, H. Guo, J. Wang, Phys. Rev. B {\bf 63}, 245407 (2001).

\bibitem{nanodcal1} Yi-Bin Hu, Yong-Hong Zhao, and Xue-Feng Wang, Front. Phys. {\bf 9}, 760 (2014).

\bibitem{nanodcal2} Wei Ji, Hong-Qi Xu, and Hong Guo, Front. Phys. {\bf 9}, 671 (2014)

\bibitem{LDAC} S. H. Wei and A. Zunger, Phys. Rev. B {\bf 57}, 8983 (1998).

\bibitem{gapexp} B. Tell and P. M. Bridenbaugh, Phys. Rev. B {\bf 12}, 3330 (1975).

\end{thebibliography}
\end{document}